\documentclass[a4paper,aps,prd,10pt,preprintnumbers,twocolumn,superscriptaddress,nofootinbib,amsmath,amssymb]{revtex4-1}
\usepackage{graphicx}
\usepackage[utf8]{inputenc}
\usepackage[T1]{fontenc}
\usepackage{cmap}
\usepackage{ulem}
\DeclareMathAlphabet{\pazocal}{OMS}{zplm}{m}{n}

\begin{document}
\title{Grey-body factors and Hawking radiation of black holes in $4D$ Einstein-Gauss-Bonnet gravity}
\author{Roman A. Konoplya} \email{roman.konoplya@gmail.com}
\affiliation{Research Centre for Theoretical Physics and Astrophysics, Institute of Physics, Silesian University in Opava, Bezručovo nám. 13, CZ-74601 Opava, Czech Republic}
\affiliation{Peoples Friendship University of Russia (RUDN University), 6 Miklukho-Maklaya Street, Moscow 117198, Russian Federation}
\author{Antonina F. Zinhailo} \email{antonina.zinhailo@physics.slu.cz}
\affiliation{Research Centre for Theoretical Physics and Astrophysics, Institute of Physics, Silesian University in Opava, Bezručovo nám. 13, CZ-74601 Opava, Czech Republic}
\begin{abstract}
The $(3+1)$-dimensional Einstein-Gauss-Bonnet theory of gravity which breaks the Lorentz invariance in a theoretically consistent and observationally viable way has been recently suggested by Aoki, Gorji and Mukohyama [arXiv:2005.03859]. Here we calculate grey-body factor for Dirac, electromagnetic and gravitational fields and estimate the intensity of Hawking radiation and lifetime for asymptotically  flat black holes in this theory. Positive coupling constant leads to much smaller evaporation rate and longer life-time of a black hole, while the negative one enhances Hawking radiation. The grey-body factors for electromagnetic and Dirac fields are smaller for larger values of the coupling constant.
\end{abstract}
\maketitle

\section{Introduction}

Higher curvature corrections added to the Einstein theory of gravity represent a broad area of alternative theories of gravity generalizing General Relativity.
Among various types of higher curvature  corrections, one of the most perspective approaches is consideration of quadratic corrections in the form of the Gauss-Bonnet term.
In $3+1$ dimensions the Gauss-Bonnet term produces the pure divergence and does not add anything to the equations of motions, so that the Einstein theory is the only metric theory of gravity which keeps diffeomorphism invariance and, at the same time, has second order equations of motion. However, a non-Lagrangian approach can be developed \cite{Glavan:2019inb,Tomozawa:2011gp}, which is based on, first, re-scaling of the coupling constant in the higher dimensional field equations and, then, taking the limit $D \rightarrow 4$. It was claimed that this way the Lovelock's theorem is bypassed \cite{Lovelock:1971yv,Lovelock:1972vz} and the Ostrogradsky instability is avoided. However, later it was shown that the above regularization scheme works only for a class of metrics for which
\begin{equation}
C^{\mu\rho\lambda\sigma}C_{\nu\rho\lambda\sigma}-\frac{1}{4}\delta^{\mu}_{\nu}C^{\tau\rho\lambda\sigma}C_{\tau\rho\lambda\sigma}=0,
\end{equation}
where $C_{\nu\rho\lambda\sigma}$ is the Weyl tensor \cite{Aoki:2020lig}. Therefore, the regularization scheme \cite{Glavan:2019inb,Tomozawa:2011gp} does not produce the four-dimensional theory of gravity \cite{Gurses:2020ofy,Arrechea:2020evj}. The well-definied theory, which, however, breaks the Lorenz-invariance and affects the dispersion relation for propagation of gravitons at high frequencies, was suggested in  \cite{Aoki:2020lig,Aoki:2020iwm}. Fortunately, the black hole solution obtained via the naive regularization  \cite{Glavan:2019inb,Tomozawa:2011gp} proved out to be an exact solution also in the full theory \cite{Aoki:2020lig,Aoki:2020iwm} as well as in the scalar-tensor theories with the Gauss-Bonnet term \cite{Lu:2020iav,Kobayashi:2020wqy,Fernandes:2020nbq,Hennigar:2020fkv}. Therefore, one can safely consider test fields in the background of these black holes as well as gravitational perturbations using the same regularization scheme, but keeping in mind that the gravitational sector may be modified in the full theory in the regime of high frequencies \cite{Aoki:2020iwm}. It is well-known that in four-dimensional spacetimes, gravitons contribute only about one-two percent into the total amount of radiation around black holes. Therefore, we can neglect radiation of gravitons when estimating the intensity of Hawking radiation and black-hole's lifetime.

Black holes in either higher- \cite{Boulware:1985wk,Wheeler,Wiltshire:1985us,Cai:2001dz} or four-dimensional  \cite{Glavan:2019inb,Konoplya:2020qqh}   Einstein-Gauss-Bonnet theories and their Lovelock generalizations are limited by strong constrains on their parameters due to the gravitational instability of black-hole spacetimes \cite{Dotti:2005sq,Gleiser:2005ra,Konoplya:2008ix,Konoplya:2017lhs,Konoplya:2017zwo,Takahashi:2010ye,Takahashi:2011qda,Gonzalez:2017gwa,Cuyubamba:2016cug}. The black holes are stable only provided the coupling constants associated with the higher curvature terms are small enough. Therefore, the allowed deviations from the Tangherlini (or Meyrs-Perry) geometry for higher curvature corrected $D>4$ black holes are relatively small, what results in a relatively small deviations of observable quantities, such as quasinormal modes \cite{Konoplya:2004xx,Abdalla:2005hu} or iron-line radiation spectra \cite{Nampalliwar:2018iru}. The same may be true for black holes in the novel $D=4$ Einstein-Gauss-Bonnet theory with the exception that the stability region for negative values of the Gauss-Bonnet coupling constant $\alpha$ is much larger than that for positive $\alpha$, as has been recently shown in \cite{Konoplya:2020bxa,Konoplya:2020qqh,Konoplya:2020der}. Still, the final solution of the black-hole stability problem in the well-defined theory  \cite{Aoki:2020lig,Aoki:2020iwm} has not been studied.

While the classical (that is, quasinormal) spectrum of higher curvature corrected black holes differs from its Einsteinian limit relatively softly \cite{Zinhailo:2019rwd,Konoplya:2019hml,Churilova:2020aca} the Hawking radiation is known to be much more sensitive characteristic when the Gauss-Bonnet coupling is turned on \cite{Konoplya:2019hml}. Moreover, it is known that even slight deformations of the Tangherlini geometry due to the Gauss-Bonnet term lead to considerable suppression of Hawking radiation by a few orders \cite{Konoplya:2010vz,Rizzo:2006uz}. Therefore, it would be interesting to learn whether similar effects occur for Hawking radiation in the $4D$-Einstein-Gauss-Bonnet theory. Even though some thermodynamical properties in this $4D$ theory have been recently investigated in a few works   \cite{HosseiniMansoori:2020yfj,Singh:2020xju,Hegde:2020xlv,Wei:2020poh}, the study of Hawking radiation has been by now limited by consideration of grey-body factors and associated power spectra for a test scalar field only  \cite{Zhang:2020qam}. Thus, no estimations for Hawking radiation of real Standard Model particles in the $4D$ -Einstein-Gauss-Bonnet theory exist so far.

Here we will compute grey-body factors and energy emission rates for electromagnetic, Dirac and gravitational fields in the vicinity of an asymptotically flat black holes in the $(3+1)$-dimensional Einstein-Gauss-Bonnet theory. This includes calculations of radiation flows for neutrinos, photons and gravitons as well as for ultra-relativistic electrons and protons. The calculations for gravitons are made here via the regularization scheme \cite{Glavan:2019inb,Tomozawa:2011gp} which is valid in scalar-tensor theories \cite{Lu:2020iav,Kobayashi:2020wqy,Fernandes:2020nbq,Hennigar:2020fkv}, because the additional (scalar) degree of freedom is not dynamical \cite{Lu:2020iav}, but may get further corrections in the high frequency regime when considering the full theory \cite{Aoki:2020lig,Aoki:2020iwm}. As the contribution of gravitons is less than only two percents, this allows us to estimate the lifetime of the black hole for various values of the Gauss-Bonnet coupling constant.

The paper is organized as follows. In sec. II we give the basic information on the $D=4$ Einstein-Gauss-Bonnet gravity and the black-hole solution. Sec. III is devoted to deduction of the wave equations for the electromagnetic, Dirac and gravitational fields. Sec. IV discusses the boundary conditions for the scattering problem used the grey-body factors calculated with the help of the WKB method. In Sec. V we calculate the energy emission rates for Hawking radiation and estimate lifetime of the black hole under consideration. Finally, in the Conclusion, we summarize the obtained results and mention a few open questions related to the Hawking radiation in this theory.

\section{$4D$ Einstein-Gauss-Bonnet theory and black-hole metric}
A crucial aspect for our consideration of Hawking radiation is that the black-hole solution obtained as a result of the dimensional regularization suggested in \cite{Glavan:2019inb}, is also an exact solution of the well-defined truly four-dimensional Aoki-Gorji-Mukohyama theory \cite{Aoki:2020lig} or  theories with extra scalar degrees of freedom \cite{Lu:2020iav,Kobayashi:2020wqy,Fernandes:2020nbq,Hennigar:2020fkv}.
Thus, in addition to the dimensional regularization, there are three approaches where the same black-hole solution appears:
\begin{enumerate}
\item a subclass of the Horndeski theory  obtained via the Kaluza-Klein reduction of a $D$-dimensional theory with a scalar field $(\partial\phi)^4$ \cite{Lu:2020iav,Kobayashi:2020wqy}.
\item a similar approach, but without any assumption on the structure of the Kaluza-Klein sector proposed in \cite{Fernandes:2020nbq,Hennigar:2020fkv}. In \cite{Aoki:2020lig} it was shown that there is an infinite coupling problem when trying to construct the consistent quantum description of these two approaches. Nevertheless, one should be able to consider these scalar-tensor theories at least in the classical limit, for example for effective description of large astrophysical black holes.
\item A consistent and full $4D$ Aoki-Gorji-Mukohyama theory \cite{Aoki:2020lig} allowing for Hamiltonian description, which uses the ADM decomposition \cite{Aoki:2020lig}.
This theory breaks the Lorenz invariance via modification of the dispersion relations in the UV regime, making the whole approach consistent with the current observations in the IR regime.
\end{enumerate}
In the all of the above approaches the exact solution, describing an asymptotically flat four-dimensional black hole with Gauss-Bonnet corrections has the form

\begin{equation}\label{sphansatz}
ds^2 = -f(r)dt^2
	+f^{-1}(r)dr^2
	+r^2d\Omega_{2}^2,
\end{equation}
where
\begin{align}\label{sch-de}
f(r) = 1 + \frac{r^2}{\alpha }
	\Biggl[ 1\pm \biggr( \!1 \!+\! \frac{4 \alpha M}{r^3} \biggr)^{\!\! 1/2 \,} \!
	\Biggr],
\nonumber
\end{align}
and $\alpha$ is the Gauss-Bonnet coupling constant.
There are two branches of solutions if~$\alpha\!>\!0$, but, if $\alpha\!<\!0$, there is no real solution for $r^3\!<\!-4 \alpha M$.  Here we will study ``the minus'' case of the above metric, as it leads to an asymptotically flat solution, unlike ``the plus'' case, which is asymptotically de Sitter one.

The event horizon is the larger root of the following ones:
\begin{equation}
r^{\ss H}_\pm = M \Biggl[ 1 \pm\sqrt{1 - \frac{\alpha}{2}} \ \Biggr] \, .
\end{equation}
Various properties of the $4D$ Einstein-Gauss-Bonnet black holes and, broader, the theory itself, have been recently studied in a number of works \cite{Guo:2020zmf,Fernandes:2020rpa,Wei:2020ght,Kumar:2020owy,Ghosh:2020vpc,Doneva:2020ped,Zhang:2020qew,Lu:2020iav,Konoplya:2020ibi,Ghosh:2020syx,Kobayashi:2020wqy,Liu:2020vkh,Kumar:2020uyz,Roy:2020dyy,Singh:2020nwo,Islam:2020xmy,Kumar:2020xvu}.
Notice also that the above introduced black-hole metric was considered earlier in a different context connected with corrections to the entropy formula  \cite{Cognola,Cai:2009ua}.

\section{Wave equation for Dirac, Maxwell and gravitational fields}

\subsection{Test fields}

The general covariant equations for electromagnetic and Dirac fields have the following form
\begin{equation}\label{Maxwell}
\frac{1}{\sqrt{-g}}\partial_\mu \left(F_{\rho\sigma}g^{\rho \nu}g^{\sigma \mu}\sqrt{-g}\right)=0\,, \quad (Maxwell)
\end{equation}
\begin{equation}\label{Dirac}
\gamma^{\alpha} \left( \frac{\partial}{\partial x^{\alpha}} - \Gamma_{\alpha} \right) \Psi=0, \quad (Dirac)
\end{equation}
where $F_{\rho\sigma}=\partial_\rho {\pazocal A}_{\sigma}-\partial_\sigma {\pazocal A}_{\rho}$ and ${\pazocal A}_\mu$ is a vector potential; $\gamma^{\alpha}$ are noncommutative gamma matrices and $\Gamma_{\alpha}$ are spin connections in the tetrad formalism  \cite{Brill:1957fx}.
After some algebra one can separate the angular variables in equations (\ref{Maxwell}, \ref{Dirac}) and rewrite the wave equation in the following general master form  \cite{Brill:1957fx,Konoplya:2011qq}:
\begin{equation}  \label{klein-Gordon}
\dfrac{d^2 \Psi}{dr_*^2}+(\omega^2-V(r))\Psi=0.
\end{equation}
where  $r_*$  is the ``tortoise coordinate''
$
dr_*= d r/f(r).
$
The effective potentials are
\begin{equation}\label{empotential}
V_{1}(r) = f(r)\frac{\ell(\ell+1)}{r^2}.
\end{equation}
\begin{equation}
V_{\pm\frac{1}{2}}(r) = \frac{k}{r}f(r) \left(\frac{k}{r}\mp\frac{\sqrt{f(r)}}{r}\pm (\sqrt{f(r)})'\right),
\end{equation}
where $\ell=1, 2, 3...$ and $k=1, 2, 3,...$ are multipole numbers.
The effective potentials for electromagnetic and ``plus'' sign Dirac fields have the form of a positive definite potential barrier with a single maximum. The effective potential for the ``minus'' sign Dirac field has a negative gap near the event horizon and it is iso-spectral to the  ``plus'' potential, what was shown for a generic spherically symmetric black holes in \cite{Zinhailo:2019rwd}. Consequently, the corresponding quasinormal spectrum of the Dirac field has no growing modes indicating any kind of instability \cite{Churilova:2020aca}.

\begin{figure*}
\includegraphics[angle=0.0,width=0.5\linewidth]{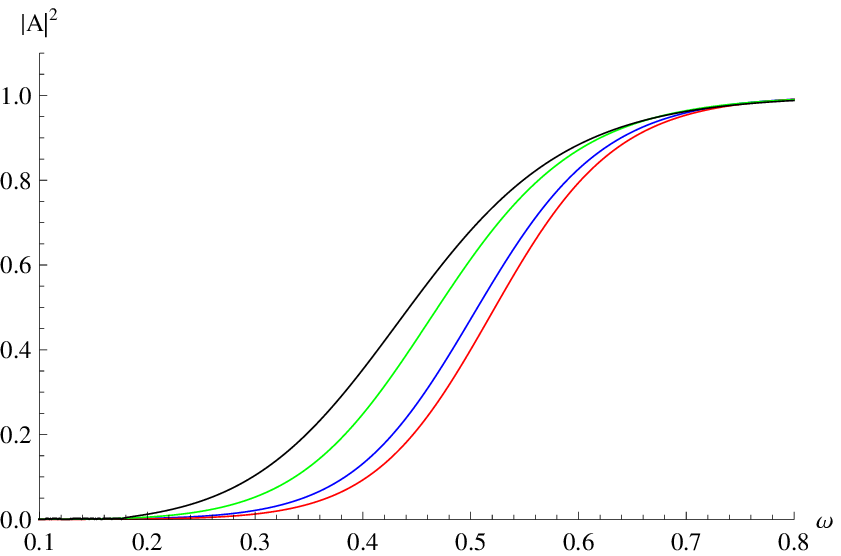}~~~~\includegraphics[angle=0.0,width=0.5\linewidth]{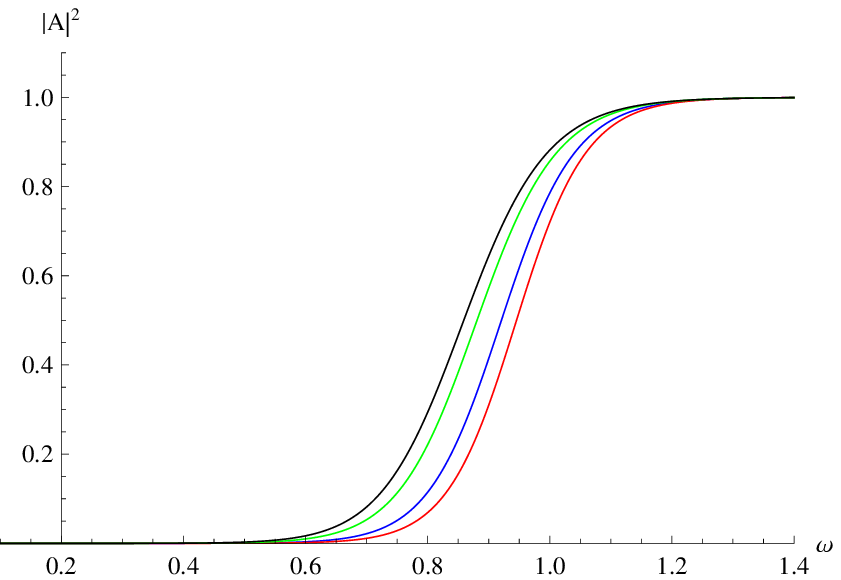}
\caption{Grey-body factors of the electromagnetic field computed with the sixth order WKB method: $M=1/2$, $\ell=1$ (left) and $\ell=2$ (right), $\alpha=0.15$ (red), $\alpha=0$ (blue), $\alpha=-0.3$ (green), $\alpha=-0.5$ (dark blue).}
\label{fig1}
\end{figure*}
\begin{figure*}
\includegraphics[angle=0.0,width=0.5\linewidth]{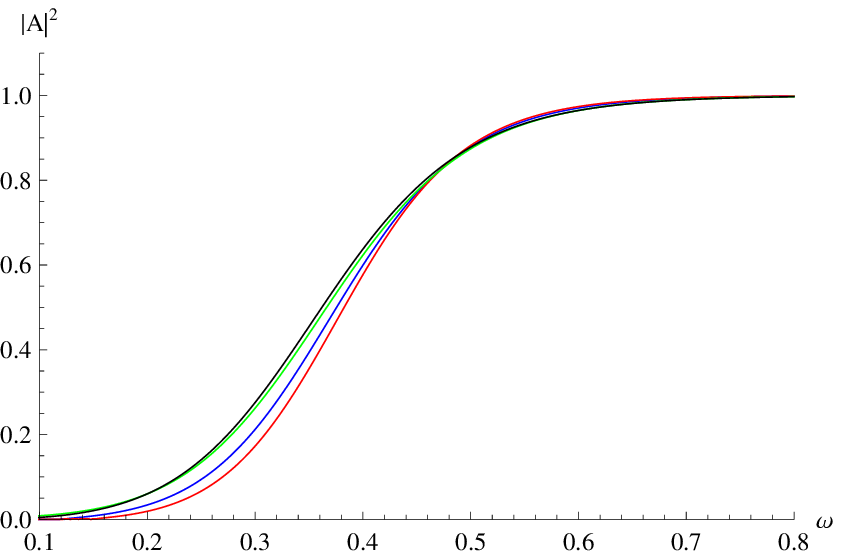}~~~~\includegraphics[angle=0.0,width=0.5\linewidth]{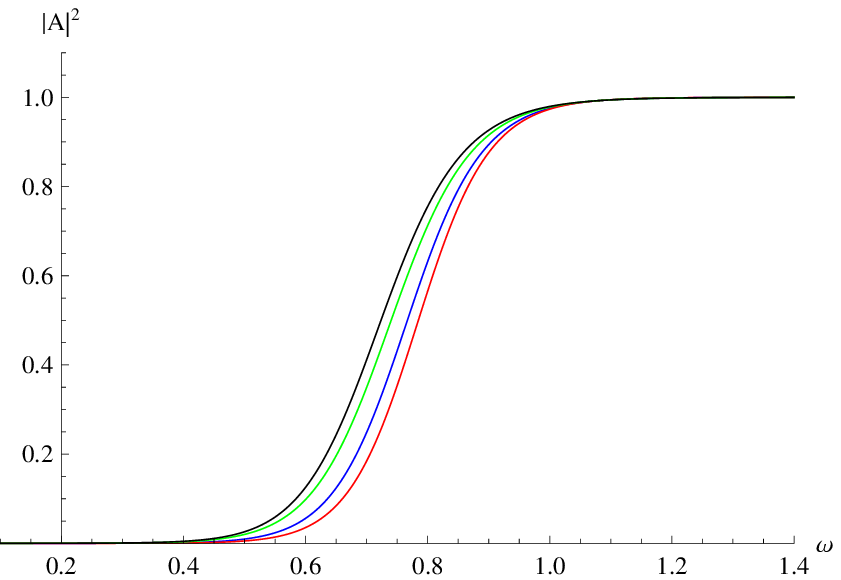}
\caption{Grey-body factors of the Dirac field computed with the fifth order WKB method: $M=1/2$, $k=1$ (left) and $k=2$ (right), $\alpha=0.15$ (red), $\alpha=0$ (blue), $\alpha=-0.3$ (green), $\alpha=-0.5$ (dark blue).}
\label{fig2}
\end{figure*}
\begin{figure*}
\includegraphics[angle=0.0,width=0.5\linewidth]{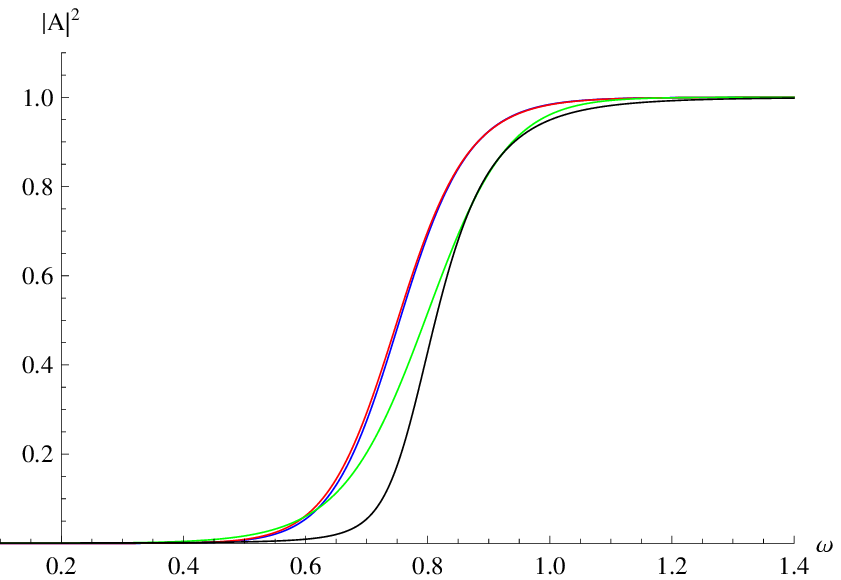}~~~~\includegraphics[angle=0.0,width=0.5\linewidth]{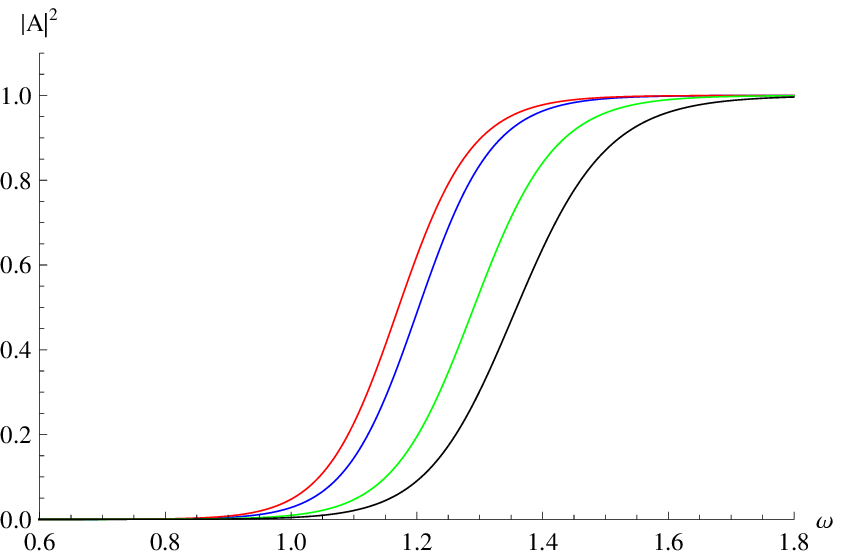}
\caption{Grey-body factors of the vector type of gravitational perturbation computed with the fourth order WKB method: $M=1/2$,  $\ell=2$ (left) and $\ell=3$ (right), $\alpha=0.15$ (red), $\alpha=0$ (blue), $\alpha=-0.3$ (green), $\alpha=-0.5$ (dark blue).}
\label{fig3}
\end{figure*}

\begin{figure*}
\includegraphics[angle=0.0,width=0.5\linewidth]{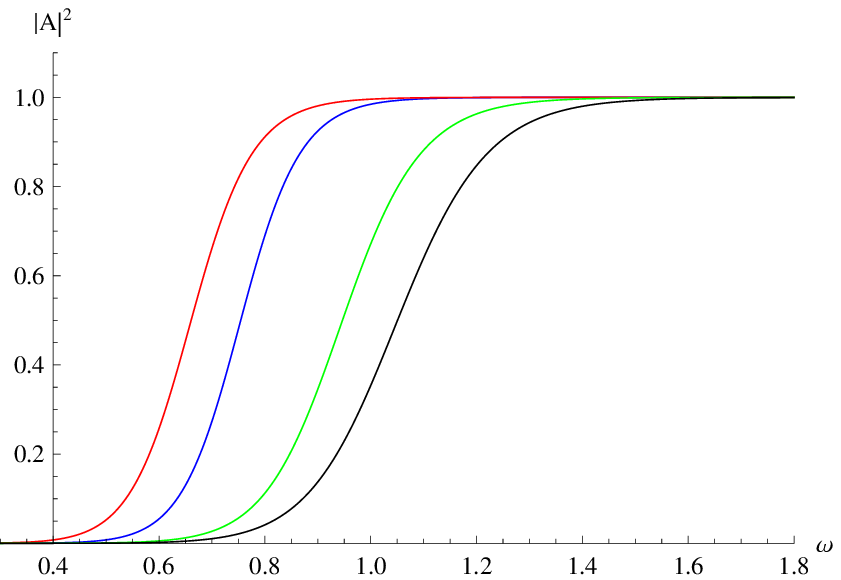}~~~~\includegraphics[angle=0.0,width=0.5\linewidth]{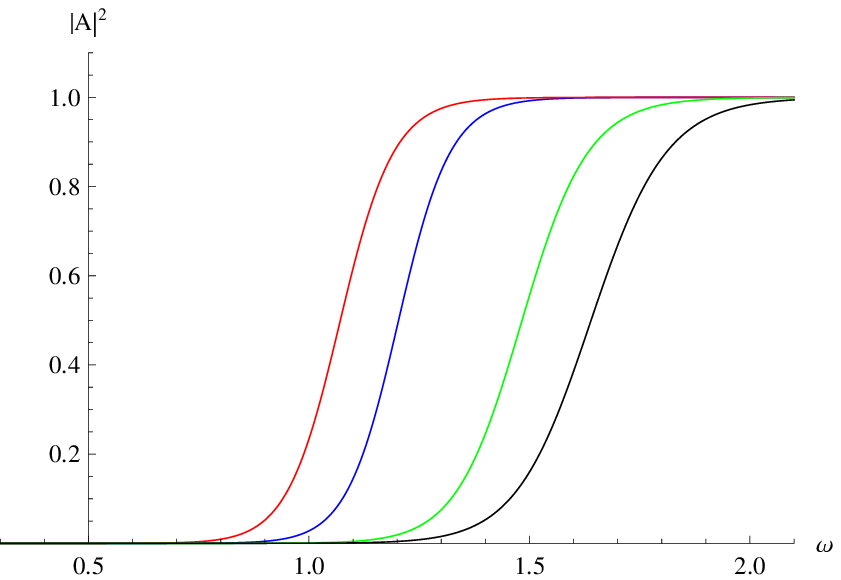}
\caption{Grey-body factors of the scalar type of gravitational perturbation computed with the fourth (for $\ell=2$) and fifth (for $\ell=3$) order WKB method: $M=1/2$,  $\ell=2$ (left) and $\ell=3$ (right), $\alpha=0.15$ (red), $\alpha=0$ (blue), $\alpha=-0.3$ (green), $\alpha=-0.5$ (dark blue).}
\label{fig4}
\end{figure*}

\subsection{Gravitational field}

In \cite{Takahashi:2010ye} it was shown that after the decoupling of angular variables and some algebra, the gravitational perturbation equations can be reduced to the second-order master differential equations.
The explicit forms of the effective potentials $V_s(r)$, $V_v(r)$ for scalar and vector types of gravitational perturbations respectively are given by
\begin{eqnarray}\nonumber
V_v(r)&=&\frac{(\ell-1)(\ell+n)f(r)T'(r)}{(n-1)rT(r)}+R(r)\frac{d^2}{dr_*^2}\Biggr(\frac{1}{R(r)}\Biggr),\\\nonumber
V_s(r)&=&\frac{2\ell(\ell+n-1)f(r)P'(r)}{nrP(r)}+\frac{P(r)}{r}\frac{d^2}{dr_*^2}\left(\frac{r}{P(r)}\right),
\end{eqnarray}
where $n=D-2$, $\ell=2,3,4,\ldots$ is the multipole number, $T(r)$ is given in \cite{Takahashi:2010ye}, and
\begin{equation}
T(r)\equiv r P'[\psi(r)]
\end{equation}
\begin{equation}
R(r)=r\sqrt{|T'(r)|},
\end{equation}
\begin{equation}
P(r)=\frac{2(\ell-1)(\ell+n)-nr^3\psi'(r)}{\sqrt{|T'(r)|}}T(r).
\end{equation}
The new function $\psi(r)$ is defined as follows:
\begin{equation}\label{Lfdef}
\psi(r) = \frac{1- f(r)}{r^2}.
\end{equation}
One can see that the vector and scalar types of gravitational perturbations (also called axial and polar ones respectively) are not iso-spectral as it takes place for the Einstein relativity. There, the corresponding effective potentials for axial and polar perturbations are related via the Darboux transformations and it is sufficient to analyze only one of the two potentials and multiply the final energy emissions by two. In our case every channel of gravitational perturbations contributes different grey-body factors.

\section{The scattering problem: boundary conditions, grey-body factors and the WKB approach}

We will study wave equation (\ref{klein-Gordon}) with the boundary conditions allowing for incoming waves from infinity. Owing to the symmetry of the scattering properties this is identical to the scattering of a wave coming from the horizon, what is natural if one wants to know the fraction of particles reflected back from the potential barrier to the horizon. The scattering boundary conditions for eq. (\ref{klein-Gordon}) have the following form
\begin{equation}\label{BC}
\begin{array}{ccll}
    \Psi &=& e^{-i\omega r_*} + R e^{i\omega r_*},& r_* \rightarrow +\infty, \\
    \Psi &=& T e^{-i\omega r_*},& r_* \rightarrow -\infty, \\
\end{array}%
\end{equation}
where $R$ and $T$ are the reflection and transmission coefficients, and we have
\begin{equation}\label{1}
\left|T\right|^2 + \left|R\right|^2 = 1.
\end{equation}
Once the reflection coefficient is calculated, we can find the transmission coefficient for each multipole number $\ell$ by the using the WKB approach:
\begin{equation}
\left|A_{\ell}\right|^2=1-\left|R_{\ell}\right|^2=\left|T_{\ell}\right|^2.
\end{equation}
\begin{equation}\label{moderate-omega-wkb}
R = (1 + e^{- 2 i \pi K})^{-\frac{1}{2}},
\end{equation}
where $K$ can be determined from the following equation:
\begin{equation}
K - i \frac{(\omega^2 - V_{0})}{\sqrt{-2 V_{0}^{\prime \prime}}} - \sum_{i=2}^{i=6} \Lambda_{i}(K) =0.
\end{equation}
Here $V_0$ is the maximum of the effective potential, $V_{0}^{\prime \prime}$ is the second derivative of the
effective potential in its maximum with respect to the tortoise coordinate $r_{*}$, and $\Lambda_i$  are higher order WKB corrections which depend on up to $2i$th order derivatives of the effective potential at its maximum \cite{Schutz:1985zz,Iyer:1986np,Konoplya:2003ii,Matyjasek:2017psv,Hatsuda:2019eoj} and $K$. This approach at the 6th WKB order was used for finding transmission/reflection coefficients of various black holes and wormholes in \cite{Konoplya:2019ppy,Konoplya:2019hml} and the comparison of the WKB results for the energy emission rate of Schwarzschild black hole done in \cite{Konoplya:2019ppy} are in excellent concordance with the numerical calculations of the well-known work by Don Page \cite{Page:1976df}. Here we will mostly use the 6th order WKB formula of \cite{Konoplya:2003ii} and, sometimes, apply lower orders when small frequencies and lower multipoles are under consideration. Fortunately, the WKB method works badly for small frequencies only, that is, in the region where the reflection is almost total and the grey-body factors are close to zero. Therefore, this inaccuracy of the WKB approach at small frequencies does not affect our estimations of the energy emission rates.
As the WKB method is very well known (see, for example reviews \cite{Konoplya:2019hlu,Konoplya:2011qq} and references therein), we will not describe it here in more detail.

\begin{figure}
\centerline{\resizebox{\linewidth}{!}{\includegraphics*{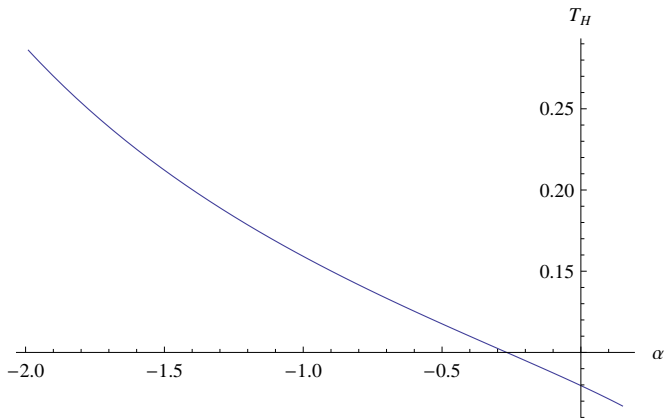}}}
\caption{Hawking temperature $T_{H}$ as a function of $\alpha$, $M=1/2$.}\label{fig5}
\end{figure}

\begin{figure*}
\includegraphics[angle=0.0,width=0.5\linewidth]{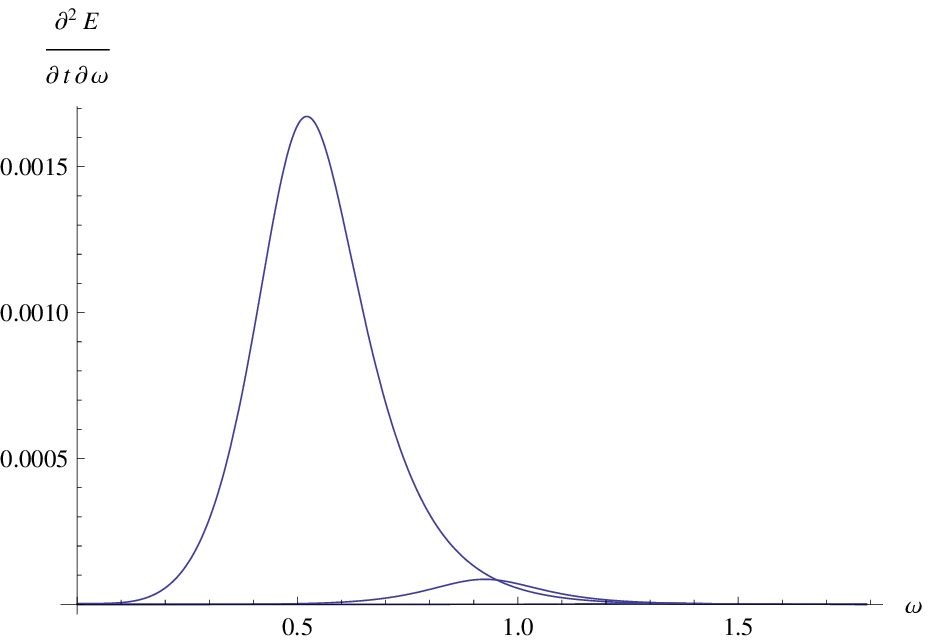}\includegraphics[angle=0.0,width=0.5\linewidth]{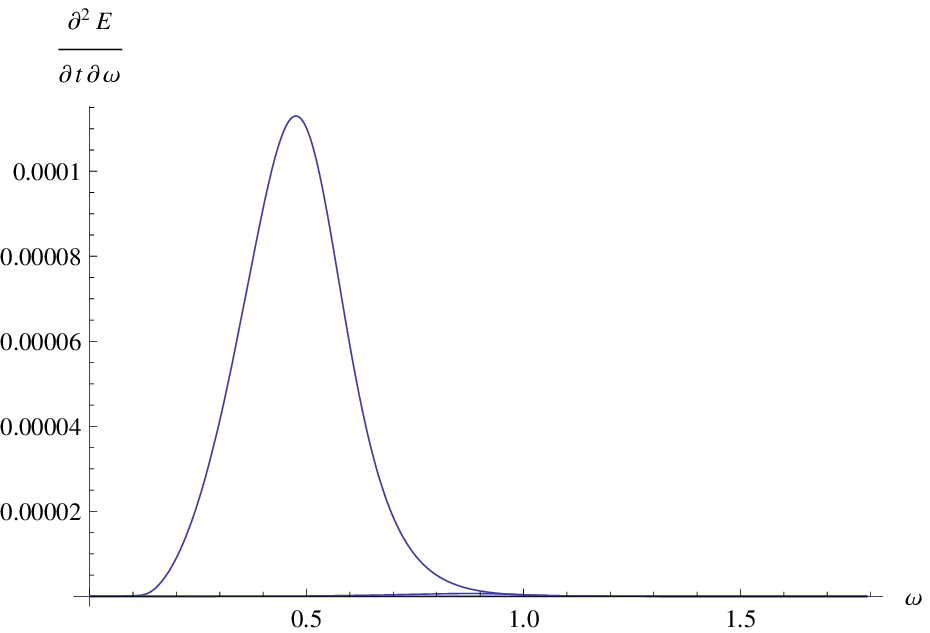}
\caption{Energy emission rate as a function of $\omega$ for the electromagnetic field for $\ell =1$ (top), $2$ and $3$ (bottom), $\alpha=-0.3$ (left), $\alpha=0.15$ (right) $M=1/2$. The contribution of the lowest multipole is dominant, $\ell=2$ only slightly corrects the total emission, while $\ell=3$ is almost negligible.}
\label{fig6}
\end{figure*}

\begin{figure}
\centerline{\resizebox{\linewidth}{!}{\includegraphics*{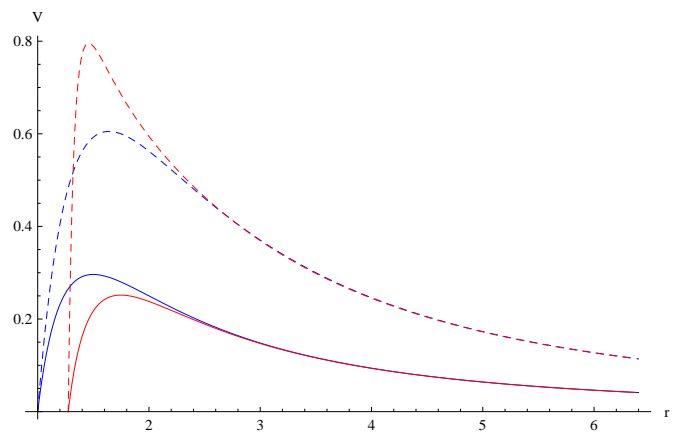}}}
\caption{Effective potentials for $\ell=1$ electromagnetic (continuous curve) and $\ell=2$ axial gravitational (dashed curve) perturbations;  $\alpha =0$ (blue) and $\alpha =-0.7$ (red).}\label{fig7}
\end{figure}

From figs. \ref{fig1}-\ref{fig4} one can see that grey-body factors of test and gravitational fields behave qualitatively differently.
For electromagnetic and Dirac fields, the transmission coefficient (given by the grey-body factor) is higher when the Gauss-Bonnet coupling $\alpha$ constant is decreasing from positive values through zero to negative values. For the axial and polar gravitational perturbations the situation is opposite: when $\alpha$ is decreased from negative to positive values, the grey-body factors are decreasing as well, leading to suppression of a fraction of gravitons penetrating the potential barrier. This effect can be easily explained from the form of the effective potentials given on fig. \ref{fig7} (continuous line): the smaller $\alpha$ is, the lower is the potential barrier for electromagnetic and Dirac fields, which allows for more particles to penetrate the potential barrier and achieve the observer detecting the incoming flow of Hawking radiation. On the contrary, the effective potential of gravitational field (dashed line on fig. \ref{fig7} )  becomes higher when $\alpha$ is decreased, what works on behave of lower grey-body factors for gravitons.

\section{Hawking radiation}

We will assume that the black hole is in the state of thermal equilibrium with its environment in the following sense: the temperature of the black hole does not change between emissions of two consequent particles. This implies that the system can be described by the canonical ensemble (see, for example, a review \cite{Kanti:2004nr}). Then, the energy emission rate for Hawking radiation is described by the well-known formula \cite{Hawking:1974sw}:
\begin{align}\label{energy-emission-rate}
\frac{\text{d}E}{\text{d} t} = \sum_{\ell}^{} N_{\ell} \left| \pazocal{A}_l \right|^2 \frac{\omega}{\exp\left(\omega/T_\text{H}\right)\pm1} \frac{\text{d} \omega}{2 \pi},
\end{align}
were $T_H$ is the Hawking temperature, $A_l$ are the grey-body factors, and $N_l$ are the multiplicities, which depend on the space-time dimension, the number of species of particles and $l$.
The Hawking temperature is given by the Hawking formula \cite{Hawking:1974sw}
$
T = f\prime(r)/(4 \pi)|_{r=r_{H}},
$
which, in our case, has the following form in the limit of small $\alpha$
\begin{equation}
T = \frac{1}{8 \pi  M}-\frac{\alpha }{32 \left(\pi  M^3\right)}-\frac{\alpha ^2}{512 \left(\pi
   M^5\right)}+O\left(\alpha ^3\right).
\end{equation}
From fig. \ref{fig5} we can see that the linear term describes the Hawking temperature quite well in the whole region of stability for positive $\alpha$.

The multiplicity factors for the four dimensional spherically symmetrical black holes case consists from
the number of degenerated $m$-modes (which are $m = -\ell, -\ell+1, ....-1, 0, 1, ...\ell$, that is  $2 \ell +1$ modes) multiplied by the number of species of particles which depends also on the number of polarizations and helicities of particles. Therefore, we have
\begin{equation}
N_{\ell} = 2 (2 \ell+1) \qquad (Maxwell),
\end{equation}
\begin{equation}
N_{\ell} = 8 k \qquad (Dirac),
\end{equation}
\begin{equation}
N_{\ell} = 2 (2 \ell+1) \qquad (gravitational).
\end{equation}
The multiplicity factor for the Dirac field is fixed taking into account both the ``plus'' and ``minus'' potentials which are related by the Darboux transformations, what leads to the iso-spectral problem \cite{Zinhailo:2019rwd} and the same grey-body factors for both chiralities. We will use here the ``minus'' potential, because the WKB results are more accurate for that case in the Schwarzschild limit.

\begin{table*}
\begin{tabular}{|c|c|c|c|c|c|c|}
  \hline
  \hline
  $\alpha$ & $dE/dt$ (Dirac) &  $dE/dt$ (Maxwell) &  Vector~grav.  &  Scalar~grav. & $\tau_{1} (M_{0}/kg)^{3} $ & $\tau_{2} (M_{0}/kg)^{3} $ \\
   \hline
 -0.3&  0.002436 & 0.000580 & 0.0000485 & 0.0000629 & $2.22 \cdot 10^{-18}$  & $1.25 \cdot 10^{-18}$ \\
    \hline
  -0.15&  0.001425 & 0.000373 & 0.0000260 & 0.0000249 & $3.76 \cdot 10^{-18}$ & $2.12 \cdot 10^{-18}$ \\
   \hline
 0& 0.000646 & 0.000137 & $7.7 \cdot 10^{-6}$  & $7.7 \cdot 10^{-6}$ & $8.7 \cdot 10^{-18}$ & $4.8 \cdot 10^{-18}$ \\
  \hline
  0.05&  0.000471 &0.000091 & $4.929 \cdot 10^{-6}$ & $6.911 \cdot 10^{-6}$ &  $1.21 \cdot 10^{-17}$
  &  $6.65 \cdot 10^{-18}$ \\
   \hline
  0.15& 0.000222 & 0.000034 & $1.55 \cdot 10^{-6}$  & $5.327 \cdot 10^{-6}$  &  $2.64 \cdot 10^{-17}$
 &  $1.43 \cdot 10^{-17}$ \\
    \hline
  \hline
\end{tabular}
\caption{Energy emission rates for Maxwell and Dirac particles (in the units $2 M=1$), lifetimes of the black hole in the regime of negligible radiation of massive particles $\tau_{1}$ and in the ultra-relativistic regime $\tau_{2}$.}\label{table2}
\end{table*}

It is well-known \cite{Page:1976df} that there are two qualitatively different regimes of particles emission. The first regime happens when the black hole mass is large enough and radiation of massive particles are negligibly small. In this regime the radiation occurs mainly due to massless electron and muon neutrinos, photons, and gravitons. When the black-hole mass $M$ is sufficiently small, emission of electrons and positrons will occur ultra-relativistically, as the wave equation for a massive field with mass $\mu$ depends on the term $\mu M$. In this ultra-relativistic regime, the law of radiation for positrons and electrons can be approximated by that for a massless Dirac field and the emission rate of all the Dirac particles must be simply doubled.

Let us assume that the peak in the Dirac particles' spectrum  $\partial^{2} E/\partial t \partial \omega$ occurs at some $\omega \approx \xi M^{-1} $, then the range of ultra-relativistic radiation of massive particles is determined as follows:
\begin{equation}\nonumber
m_{e} = 4.19 \times 10^{-23} m_{p} \ll \xi M^{-1} \ll m_{\mu} = 8.65 \times 10^{-21} m_{p}.
\end{equation}
This inequality can be rewritten as follows:
\begin{equation}
\xi^{-1} \cdot 10^{11} kg. \ll M \ll \xi^{-1} \cdot 2 \times 10^{12} kg.
\end{equation}
As can be seen from fig. \ref{fig6}, for the example of an electromagnetic field, the maximum of the spectrum shifts towards larger $\omega M$ when $\alpha$ is increased very slightly, so that it will not influence any our further estimations.

The energy emitted causes the black-hole mass to decrease at the following rate \cite{Page:1976df}
\begin{equation}
\frac{d M}{d t} = -\frac{\hbar c^4}{G^2} \frac{\alpha_{0}}{M^2},
\end{equation}
where we have restored the dimensional constants. Here $\alpha_{0} = d E/d t$ is taken for a given initial mass $M_{0}$. Since most of its time the black hole spends near  its original state $M_{0}$ and integrating of the above equation  gives us the life-time of a black hole:
\begin{equation}
\tau = \frac{G^2}{\hbar c^4} \frac{M_{0}^3}{3 \alpha_{0}}.
\end{equation}
Here $\alpha_{0}$ is the energy emission rate  that can be calculated as a sum over all the fields in Table I for non-ultrarelativistic regime and with the double weight of Dirac particles in the ultra-relativistic regime. The utlra-relativistic lifetimes are given in parentheses in Table I.

From Table I we can see that the Hawking radiation is enhanced for all fields when the coupling constant $\alpha$ is decreasing, so that negative values of $\alpha$ correspond to much more intensive Hawking radiation than positive ones. There two factors for this behavior. The first, and the dominant one, is the increasing of Hawking temperature when the coupling $\alpha$ is decreased: hotter black holes naturally emit particles more intensively. The effective potential of electromagnetic and Dirac fields also works for this tendency as it becomes lower for smaller $\alpha$, so that less particles are reflected back to the horizon for smaller $\alpha$.

There is one distinction from the Schwarzschild case is in the emission of gravitons via axial and polar channels: when $\alpha =0$ both channels are iso-spectral, because the corresponding effective potentials are related by the Darboux transformations. This is not so when the Gauss-Bonnet coupling is turned on. From Table I we can see that the difference in the energy emission rates along axial and polar channels are quite different and the polar channel always radiates more gravitons than the axial one. For example, for $\alpha =0.15$, which is near the threshold of instability \cite{Konoplya:2020bxa}, the energy emission rate via the polar channel is three times bigger than that for the axial one.

The total emission of gravitons is increasing when the value of the coupling constant is decreased from its positive values to negative ones. At the same time we have noticed here earlier that the effective potentials for negative $\alpha$ are higher what should suppress the fraction of gravitons penetrating the potential barrier and achieving the observer. However, the exponential factor of the temperature is apparently the dominant here over the linear contribution of grey-body factors, so that the total energy emission rate is anyway higher for smaller $\alpha$.

\section{Conclusions}

Here with the help of the higher order WKB approach we calculated grey-body factors and the corresponding energy emission rates for Dirac, electromagnetic and gravitational fields in the $4D$ Einstein-Gauss-Bonnet theory. We also estimated lifetime of the black hole for various values of the coupling constant.
We have shown that the positive coupling constant leads to considerable suppression of Hawking evaporation, while the negative one enhances it. Emission rates via the two channels of gravitational perturbations are not the same anymore: Gravitons emission via the polar channel is much higher than that through the axial one. The grey-body factors of test fields decrease when the positive coupling constant is turned on, what, together with the decreasing temperature, works for suppression of intensity of the emission.

In order to find energy emission rate of Hawking radiation at large negative $\alpha$ the WKB method does not provide sufficient accuracy in that regime. However, our main purpose here was the case of positive coupling constant for which the solution exists in the whole space and is not terminated at some finite value of the radial coordinate $r$. Numerical integration of the wave equations allows one to find grey-body factors for any large negative values of the coupling constant, which, we hope, could be the subject of future investigations. Adding such a factor as an electric charge could also be interesting to complete the picture of the black hole evaporation.

While consideration of test fields is valid in all of the above approaches (because for this one uses only the form of the background metric) the gravitational perturbations,  which we treat here via the dimensional regularization scheme, must be valid only in the scalar-tensor theory, because there the scalar field is not dynamical \cite{Lu:2020iav}. Therefore,  implying the correct quantum description of black holes, we certainly should interpret carefully the data on gravitons' emission, as in the ultra-violet regime, that is exactly when the black hole is small and intensively evaporating, the gravitons' spectrum will be corrected in the Aoki-Gorji-Mukohyama theory \cite{Aoki:2020lig}. As the contribution of gravitons is almost negligibly small for four-dimensional black holes, even the estimated emission of test fields only gives a clear picture of black-hole evaporation. Nevertheless, it is necessary also to fulfill the reduction of perturbation equations to the master wave-like form in the full well defined theory \cite{Aoki:2020lig} and, in a similar fashion, to estimate the intensity of emission of gravitons, which may be different due to the modification of the dispersion relation.

\acknowledgments{
The authors acknowledge 19-03950S GAČR grant. R. K. also  acknowledges the ``RUDN University Program 5-100''. A. F. Z. thanks the SU grant SGS/12/2019.}

\end{document}